\documentclass[a4paper,12pt]{spieman}
\usepackage{amsmath,amsfonts,amssymb}
\usepackage{graphicx}
\usepackage{setspace}
\usepackage{tocloft}
\usepackage{afterpage}
\usepackage{xcolor}

\title{Progress towards a microchannel plate detector with AlGaN photocathode and cross-strip anode for ultraviolet astronomy}

\author[a,*]{S.~Diebold}
\author[a]{J.~Barnstedt}
\author[a]{L.~Conti}
\author[b]{H.R.~Elsener}
\author[a]{L.~Hanke} 
\author[a]{M.~Höltzli}
\author[a]{C.~Kalkuhl}
\author[c]{D.~Rau} 
\author[c]{D.~Schaadt}
\author[a]{T.~Schanz}
\author[a]{B.~Stelzer}
\author[a]{K.~Werner}

\affil[a]{Institut für Astronomie und Astrophysik, Eberhard Karls Universität Tübingen, Sand 1, 72076 Tübingen, Germany}
\affil[b]{Empa, Swiss Federal Laboratories for Materials Science and Technology, Ueberlandstrasse~129, 8600 Dübendorf, Switzerland}
\affil[c]{Institut für Energieforschung und Physikalische Technologien, Technische Universität Clausthal, Leibnizstraße 4, 38678 Clausthal-Zellerfeld, Germany}

\cftpagenumbersoff{figure}
\cftpagenumbersoff{table} 
\begin{document}
\maketitle

\begin{abstract}
Microchannel plates (MCPs) were the driving detector technology for ultraviolet (UV) astronomy over many years, and still today MCP-based detectors are the baseline for several planned UV instruments. The development of advanced MCP detectors is ongoing and pursues the major goals of maximizing sensitivity, resolution, and lifetime, while at the same time decreasing weight, volume, and power consumption.

Development efforts for an MCP-based detector system for the UV are running at IAAT at the University of Tübingen. In this publication, we present our latest results towards coating aluminum gallium nitride (AlGaN) photocathodes directly on MCPs, to improve quantum detection efficiency in the far- and extreme-UV. Furthermore, we report on the implementation of a non-iterative centroiding algorithm for our coplanar cross-strip anode directly in an FPGA.
\end{abstract}


{\noindent \footnotesize\textbf{*} Address all correspondance to Sebastian Diebold, \linkable{diebold@astro.uni-tuebingen.de} }

\section{Introduction}
\label{sect:intro}

\begin{figure} [ht]
   \begin{center}
        \includegraphics[width=.6\textwidth]{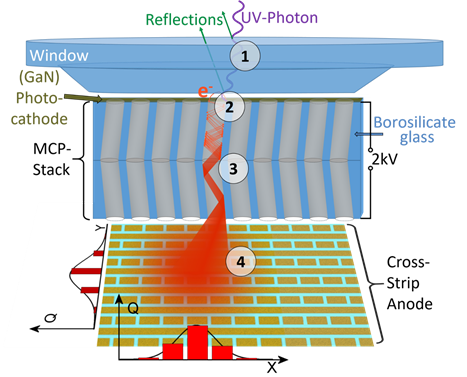}
    \end{center}
    \caption
   { \label{fig:MCP_scheme} 
Sketch of the MCP detector principle for the UV: The window (1) is optional and only applicable for wavelengths larger than about 118\,nm. Incoming UV photons hit a large-bandgap photocathode and release photoelectrons (2). These photoelectrons are accelerated in the channels of an MCP stack (3) and generate a charge cloud. The position of the charge cloud exiting the MCP stack is determined with a position-sensitive anode (4).}
\end{figure}

The principle of a microchannel plate (MCP) detector for the ultraviolet (UV) is illustrated in Fig.~\ref{fig:MCP_scheme}. UV photons pass through an entrance window and hit a photocathode with a sufficiently large bandgap to avoid sensitivity in the visible band. For wavelengths below 118\,nm no stable window material exists, thus, an open-face detector design is necessary with the photocathode deposited on top of the MCP stack. If a UV photon hits the photocathode, a photoelectron is produced via the photoelectric effect and accelerated in the microscopic channels of the MCP stack by the applied high voltage. Within these channels secondary electrons are produced in an avalanche-like process when electrons impact the walls. The channels are tilted slightly from the surface normal by a bias angle, and usually two to three MCPs are stacked with alternating bias angle direction. Besides increasing the gain and optimizing the pulse height distribution (PHD), stacking avoids spurious signals from ions that are created within the channels and move backwards to the entrance side. The typical gain of a stack of two MCPs is of the order of 10$^6$ to 10$^7$ with a peaked PHD. The exiting charge cloud is detected on a position-sensitive anode and its center corresponds to the position of the initial photon event.

The main assets of MCP detectors for the UV are their capability for single-photon counting and their visible-blindness by design, i.e.\@ a complete insensitivity to visible light. Both simplify instrument development as well as mission design, because photon counting with high time resolution can be used to correct for spacecraft and pointing jitter, and visible-blindness reduces stray light issues considerably.
Low dark count rates of 0.04\,events\,cm$^{-2}$\,s$^{-1}$ can be achieved with MCP detectors that combine novel borosilicate-glass-based MCPs with ultra-wide bandgap photocathodes such as Al$_{x}$Ga$_{1-x}$N \cite{Ertley2018}. The low power dissipation and no need for cooling further facilitate the implementation of an MCP detector in a UV instrument. An additional asset compared to other detector technologies -- of particular importance for smaller missions -- are the comparatively low costs of MCP detectors.

The details about our MCP detector design can be found in references \cite{Conti2018, Conti2022, Diebold2024}. While we successfully produced a sealed detector with an MgF$_2$ window for UV wavelengths down to about 118\,nm already some years ago\cite{Conti2022}, our efforts to coat Al$_{x}$Ga$_{1-x}$N directly on an MCP for an open-face detector are ongoing. Our long-term goal for this development lies in reaching a quantum detection efficiency in the far and extreme UV (FUV, 90--180\,nm; EUV, 10--90\,nm) that exceeds state-of-the-art photocathodes such as KBr and CsI. For efficient photoelectron emission, the creation of a negative electron affinity (NEA) surface is essential. In Mg-doped Al$_{x}$Ga$_{1-x}$N this is achieved by applying thin Cs/O layers to the surface. Previous results for our activation process are outlined in Conti et al.\@ (2022)\cite{Conti2022}.

Our detector approach aims on enabling future instruments to deliver high performance in a compact package and cover the full FUV band including the Lyman UV (LyUV, 90\,nm -- 122\,nm) as needed for the proposed TINI, CAFE, and LyRIC missions\cite{Diebold2022a, Ji2020}. Furthermore, the goal is to extend the wavelength range of our detector system down into the EUV, which is mainly driven by the planned EUV mission SIRIUS that was selected by ESA as backup in the fast-mission calls for F1 and F2 of the Cosmic Vision program and subsequently funded in UK for a phase-0 study\cite{Diebold2024, Barstow2014}. However, the future application of MCP detectors in UV astronomy is not limited to small and medium-class missions. Also for NASA's planned flagship mission Habitable Worlds Observatory (HWO) MCP detectors are the baseline technology for its shortest wavelength band 100--200\,nm \cite{Scowen2025,Curtis2025}.

In Section~\ref{sect:gan} of this publication we report on our latest results towards coating a highly efficient Al$_{x}$Ga$_{1-x}$N photocathode directly on MCPs. We also continue to improve the readout electronics for our cross-strip anode (CSA) and, hence, present in Section~\ref{sect:centroiding} our latest approach for implementing a non-iterative centroiding algorithm directly in a field-programmable gate array (FPGA).


\section{AlGaN Photocathode}
\label{sect:gan}
III-nitrides like gallium nitride (GaN), aluminum nitride (AlN) and their ternary compound alu\-mi\-num-gallium nitride (Al$_{x}$Ga$_{1-x}$N) show promising properties regarding many complex optoelectronic applications such as solar cells, high power electronics, and UV detectors used in space research \cite{Zhou17}. By varying the Al fraction $x$, the bandbap of Al$_{x}$Ga$_{1-x}$N -- and therefore the low energy cut-off in our detector application -- is tunable between 3.4\,eV ($x=0$, pure GaN) and 6.2\,eV ($x=1$, pure AlN) \cite{Muth1999}. In thin film research, III-nitrides occur in three main crystal structures: the hexagonal wurtzite structure (alpha), the cubic zinc-blende structure (beta), and the cubic rocksalt structure (gamma). While the hexagonal structure is the thermodynamically stable form, it suffers from an inherent polarity and spontaneous polarization which can negatively affect the optical and electrical properties of the material \cite{Bayr17,Park00}. The metastable cubic forms are nonpolar, therefore avoid polarization-induced fields that reduce carrier lifetimes and are thus currently gaining interest in semiconductor research.
These film structures are influenced by the underlying lattice structure of the substrate, therefore the metastable phases can be stabilized\cite{Fu12}. Thus, it is assumed that the cubic structure of the MgO substrate may stabilize the cubic forms in III-nitride films. While the beta structure has successfully been stabilized using cubic buffer layers \cite{Kaku13}, property predictions of the gamma structure rely on calculations \cite{Zhan07}. These calculations, however, show very promising optoelectrical properties for the gamma structure that may be superior to those of the hexagonal (alpha) and cubic zinc-blende (beta) structure \cite{Zhan07}. However, the most commonly used substrate materials, like (0 0 0 1) $c$-plane sapphire, have a hexagonal structure which further promotes the growth of the thermodynamically stable alpha III-nitride structure while presenting with a high lattice mismatch of around 16\,\% \cite{Melton1969}. In thin films, a high lattice mismatch can cause stress and strain in the film, resulting in a high defect density \cite{Melton1969}. For optoelectronic applications, lattice defects present as recombination sites resulting in deteriorating electrical and optical properties especially the quantum efficiency of the photocathode\cite{Jana2014}. Thus, a goal is the development of a pure-phase film with minimal lattice mismatch to the substrate.

\subsection{Growth and Characterization}
\label{ssec:methods}
We grow Al$_{x}$Ga$_{1-x}$N films with compositions $x$ ranging from 0 to 1 on square, single crystal 10~$\times$~10\,mm$^2$ MgO (1 0 0) substrates using a RIBER Compact 21 molecular beam epitaxy (MBE) system. Before growth, the substrates are heated to 130\,°C for 30 minutes in the loading chamber. In the growth chamber, they are then heated to 650\,°C, i.e. 50\,K above the growth temperature, for cleaning. The films are grown at 600\,°C over a period of 60 minutes. Reactive nitrogen flow is provided by a radiofrequency (RF) plasma cell with a forward power of 400\,W and a constant gas flow of 0.4\,sccm. Due to its poor efficiency only about 1\,\% of the nitrogen molecules are split into atoms, therefore the N-flux has to be about 100 times higher than the metal-flux for a stochimoetric ratio. The ratio of the group-III-flux (Al+Ga) to the N-flux is metal-rich at 2.9\,\% for the $Al_{0.00}Ga_{1.00}N$ and $Al_{1.00}Ga_{0.00}N$ samples and 3.7\,\% for all compound samples. The Ga- and Al-fluxes are varied for each sample to achieve Al fractions of $x = 0.00$, $0.25$, $0.50$, $0.75$, and $1.00$, respectively. The growth rate is kept constant at 0.12\,--\,0.13\,ML/s (monolayers per second).

After the growth, samples from the previous results (see Sect.~\ref{ssec:preres}) are analyzed using optical, scanning electron, and atomic force microscopy to characterize the surface. High-resolution X-ray diffraction (HRXRD) measurements are carried out using a Bruker D8 Discover system to determine the crystal structure. Lattice constants as well as layer stress and strain are investigated using $2\Theta - \omega$ scans and reciprocal space maps (RSMs).

For the latest results (see Sect.~\ref{ssec:curres}), transmission and reflection measurements are performed using a two-beam spectroscope (SPECORD 200 system by Analytik Jena). These measurements cover the wavelength range from 190\,nm to 1100\,nm in 5\,nm increments to study the films' optical behavior across the UV–Vis–NIR spectrum and to determine the bandgap of each sample.

\subsection{Previous Results}
\label{ssec:preres}
\begin{figure} [ht]
   \begin{center}
        \includegraphics[width=.8\textwidth]{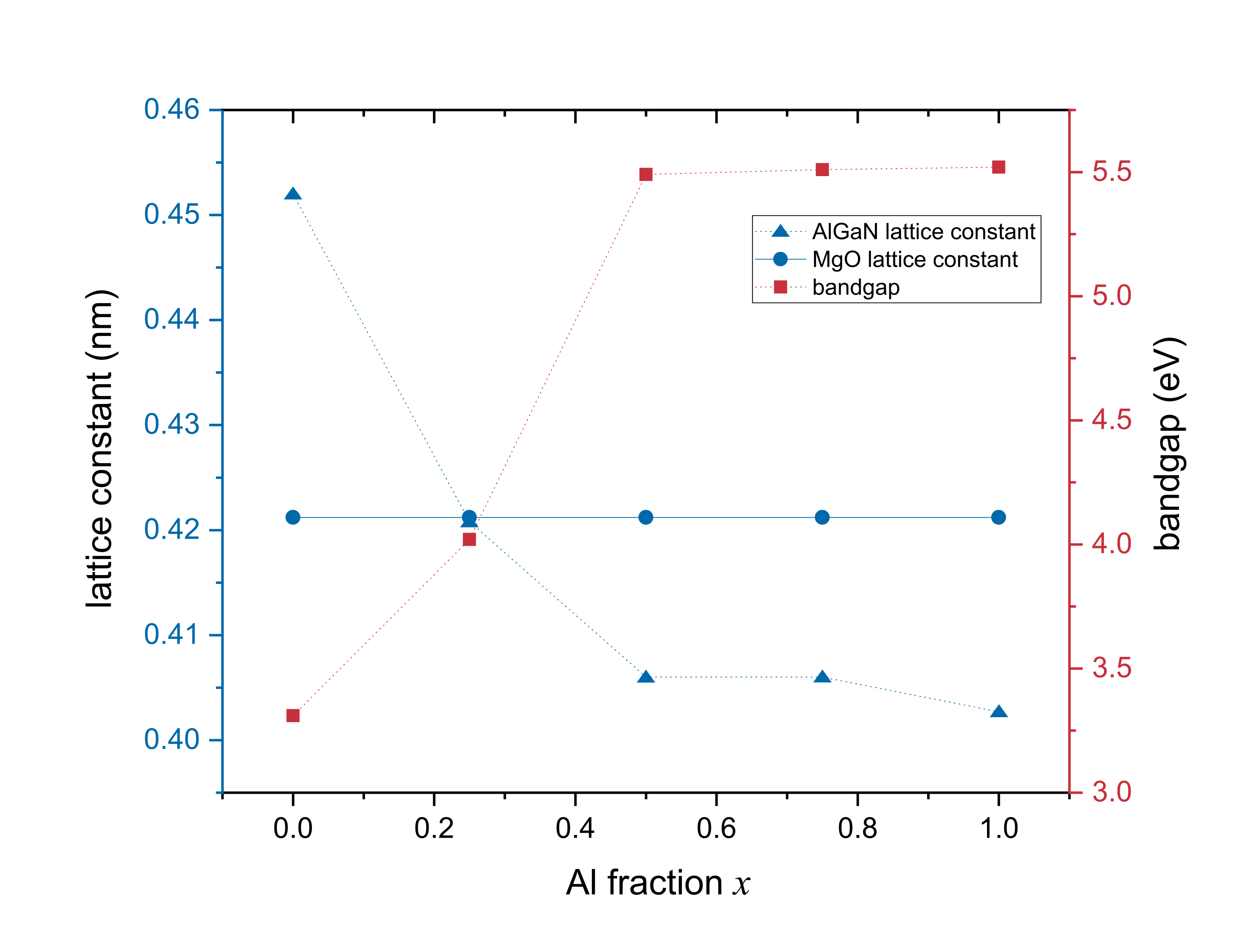}
    \end{center}
    \caption
   { \label{fig:lattice+bandgap} 
Plot of the lattice constants of MgO substrate and Al$_{x}$Ga$_{1-x}$N films for different Al fractions $x$. The lowest lattice mismatch was found at $x=0.25$. For small values of $x$ a strong dependence of the measured bandgap on the Al fraction is visible while for $x>0.5$ the bandgap is almost stable.}
\end{figure}

In previous HRXRD measurements, all samples were found to contain cubic phases. While the Al$_{1.00}$Ga$_{0.00}$N sample showed small amounts of hexagonal phases, all other samples contained pure cubic phases. The Al$_{0.00}$Ga$_{1.00}$N sample showed a beta-GaN film, while the Al$_{1.00}$Ga$_{0.00}$N sample has a gamma structure in the cubic phase. The lattice constants were determined for all samples from the $2\Theta - \omega$ scans and RSMs. These lattice constants are summarized in Table~\ref{tab:latticeconst} and illustrated in Fig.~\ref{fig:lattice+bandgap}. The lowest lattice mismatch measured was about 0.1\,\% for the Al$_{0.25}$Ga$_{0.75}$N sample. However, using Vegard's Law \cite{Vega21} a lattice match was expected at a composition around $x=0.68$, thus, far off the experimental result. Optical surface investigations revealed significant droplet formation, which Auger measurements confirmed as pure metallic Ga. This led to the assumption that the actual Al-content varies from the Al:Ga flux ratios during the growth which is assumed to be partially responsible for the deviation of the expected lattice match composition. Current research is focused on preventing droplet formation by investigating pulsed growth techniques and varying growth parameters, such as the III-to-V ratio and growth temperature.

\begin{table}[ht]
    \caption{Lattice constants of MgO substrate and Al$_{x}$Ga$_{1-x}$N films for different Al fractions $x$. The lowest lattice mismatch of about 0.1\,\% was found at $x=0.25$. For small values of $x$ a strong dependence of the measured bandgap on the Al fraction is visible, while for $x>0.5$ the bandgap is almost stable at about 5.5\,eV.}
    \label{tab:latticeconst}
    \begin{center}
        \begin{tabular}{|c|c|c|c|}
            \hline
            \rule[-1ex]{0pt}{3.5ex}  Al fraction $x$ & Al$_{x}$Ga$_{1-x}$N lattice constant & MgO lattice constant & bandgap \\
            \rule[-1ex]{0pt}{3.5ex}   & (nm) & (nm) & (eV) \\
            \hline\hline
            \rule[-1ex]{0pt}{3.5ex}  0.00 & 0.4520 & 0.4212 & 3.31 \\
            \hline
            \rule[-1ex]{0pt}{3.5ex}  0.25 & 0.4208 & 0.4212 & 4.02 \\
            \hline
            \rule[-1ex]{0pt}{3.5ex}  0.50 & 0.4060 & 0.4212 & 5.49 \\
            \hline
            \rule[-1ex]{0pt}{3.5ex}  0.75 & 0.4060 & 0.4212 & 5.51 \\
            \hline
            \rule[-1ex]{0pt}{3.5ex}  1.00 & 0.4027 & 0.4212 & 5.52 \\
            \hline
        \end{tabular}
    \end{center}
\end{table} 

\subsection{Latest Results}
\label{ssec:curres}
As previously mentioned, an advantage of the III-nitride continuous alloy system is the ability to adjust the bandgap cut-off wavelength -- so-called bandgap engineering -- by adjusting the composition\cite{Nepal2005}. This can be used to further investigate the actual Al fraction in the Al$_{x}$Ga$_{1-x}$N films with UV-Vis measurements. Each sample is first measured with the reference slot of the two-beam spectroscope left empty. These measurements correspond to the whole sample complex of film and substrate. Measurements in this setup are evaluated using the Tauc Plot Method with a baseline correction \cite{Maku18}. Subsequently, the samples are measured with an MgO substrate in the reference slot. This second step isolates the optical properties of the film itself, and the results are evaluated using the Tauc Plot Method without baseline correction, as the presence of the substrate in the reference beam makes it unnecessary.

For both measurement setups, the obtained bandgap values are identical and are summarized in Table~\ref{tab:latticeconst} and illustrated in Fig.~\ref{fig:lattice+bandgap}. Measurements using the MgO reference, isolating the Al$_{x}$Ga$_{1-x}$N compound films, confirm the visible-blindness of these films. While the determined bandgaps coincide well with the calculated lattice constants, the bandgaps significantly deviate from the expected values predicted using Vegard's Law based on the flux ratios. However, Vegard's Law can be used to approximate the actual composition of the films using the determined bandgaps, though this approach presents some challenges. It is well-known that the linear approximation inherent in Vegard's Law deviates from experimentally observed behavior. For more precise calculations, Vegard's Law needs to be augmented by introducing the bowing factor \cite{Adac85}. This factor accounts for the finite curvature observed experimentally. However, the precise value of the bowing factor for III–V compound semiconductors, especially Al$_{x}$Ga$_{1-x}$N, remains widely disputed \cite{Tsai2019, Kanoun2005}. While predictions using first principles calculations exist for the beta phase, to our knowledge no predictions for the gamma phase are available due to its rare occurrence. Thus, in this work, we limit ourselves to using linear Vegard's Law as a rough approximation. Another challenge is posed by the unknown crystal structure of the compound phases. Since the bandgap of the film strongly depends on its crystal structure, this significantly influences the results. Thus, calculations were performed considering all three possible crystal structures (alpha, beta, and gamma). Furthermore, only a range of plausible compositions will be provided. Moreover, these calculations using Vegard's Law allow for improved estimation of the crystal structure present in the compound films. Hexagonal alpha structure of the compound phases can already be ruled out based on HRXRD measurements. The calculated lattice constants indicate that if the compound films had a hexagonal structure, significant stress would be present in the films. This stress would occur regardless of crystal orientation. Since no stress was observed in HRXRD RSMs, the compound films are assumed to form in a cubic structure. Vegard's Law calculations rule out the cubic beta structure for the Al$_{0.50}$Ga$_{0.50}$N and Al$_{0.75}$Ga$_{0.25}$N films. According to these calculations, the beta structure would require a composition of $x>1$. Thus, it is concluded that the gamma phase formed. In this case, the actual Al-content of both samples is between 82\,\% and 93\,\%, which is consistent with the measured lattice constants. For the Al$_{0.25}$Ga$_{0.75}$N sample, no definite statement about the lattice structure can be made. Calculations using both cubic structures indicate that the actual Al-content is between 36\,\% and 55\,\%. Further investigation into the actual Al-content within the film lattice will be performed using XPS (X-ray photoelectron spectroscopy) measurements which are not yet available. Gonzales et al.\@ (2007) found that in compounds with different pure phase crystal structures, the preferred structure changes abruptly at a certain composition\cite{Gonz07}. It is assumed that the same process takes place in our current samples. However, the exact composition at which this change happens has not yet been determined. Further investigations are planned to clarify this phenomenon.

The results show that the metastable cubic gamma form can be stabilized within Al$_{x}$Ga$_{1-x}$N compound phases of more than 82\,\% Al-content on an MgO substrate. In current research the stabilized gamma phase for III-nitrides is a rare occurrence. However, assumable this is partly due to a lack of gamma structured substrate materials. Following the findings of Pankov et al.\@ (2002) \cite{Pank02} and Fu et al.\@ (2012) \cite{Fu12}, who both showed the influence of the substrate material on the formation of gamma-AlN, it is assumed that the gamma structure of the MgO stabilizes the gamma crystal structure of the compound films. Ongoing work focuses on p-doping of Al$_{x}$Ga$_{1-x}$N and on the suitability of ALD (atomic layer deposition) -deposited MgO layers for epitaxial Al$_{x}$Ga$_{1-x}$N growth. While standard ALD coatings on borosilicate MCPs are initially amorphous, incorporating tailored buffer layers has been shown to promote crystallinity and provide an epitaxial surface for Al$_{x}$Ga$_{1-x}$N growth. Strategies for optimizing these layered structures will be presented in a future publication.


\section{Implementing a Non-iterative Centroiding Algorithm for a Cross-strip Anode}
\label{sect:centroiding}

\subsection{The Coplanar Cross-strip Anode}
\label{ssec:XSA}
A position-sensitive readout of the charge cloud in an MCP detector is possible with different types of pattern anodes \cite{Diebold2022,Lapington2002}, or even a pixelated silicon sensor \cite{Tremsin2020}. Among the pattern anodes, selecting a cross-strip anode (CSA) avoids distortions and, thus, simplifies calibration and data analysis, while at the same time providing high resolution at comparably low gain of the order of 10$^6$, which in turn enhances the detector lifetime \cite{Siegmund2003}. The drawback of a CSA is the large number of channels that require a dedicated readout ASIC (application specific integrated circuit) for low-power consumption and a compact design. 

\begin{figure} [ht]
  \begin{center}
    \includegraphics[width=\textwidth]{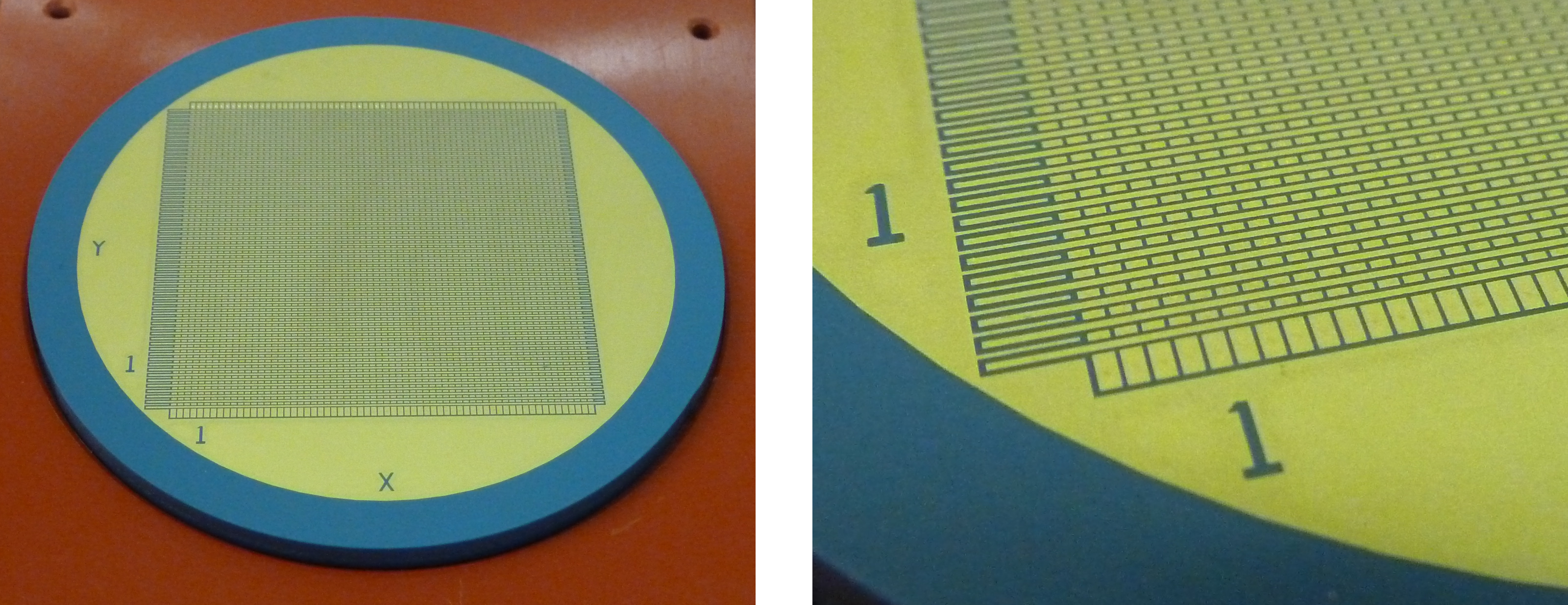}
   \end{center}
  \caption
  { \label{fig:CSA} 
Photographs of the IAAT coplanar cross-strip anode produced in an LTCC process. Left: Complete anode with gold electrodes on a blue LTCC substrate. The diameter of the substrate is about 8.8\,cm. Right: Close-up of the electrode structures. While the strips that code the y-position are complete lines, the x-strips are interrupted and form small rectangles that are connected within the substrate.}
\end{figure}

The CSA developed at IAAT features 128 strips, 64 in x- and 64 in y-direction over an area of $39\times39\,\mathrm{mm^2}$. It is manufactured in an LTCC (low-temperature cofired ceramics) process. Photographs of such an anode are shown in Fig.~\ref{fig:CSA}. While the strips that code the y-position are complete lines that are fully exposed, the strips for the x-direction consist of rectangular pads that are connected within the LTCC substrate. The width of these pads is such that the total area exposed to the charge is the same for x- and y-strips. In this coplanar design both sorts of electrodes lie within the same plane, which reduces cross-talk and increases the manufacturing yield as shortages between strips are avoided.

\subsection{A Non-iterative Centroiding Algorithm}
\label{ssec:algo}
As shown in Tremsin et al.\@ (1999) the cross-sectional shape of an electron cloud leaving an MCP stack can be approximated well by a Gaussian distribution\cite{Tremsin1999}, although in the direction of the bias angle of the MCP pores an asymmetry is present that leads to a slightly elliptic footprint\cite{Tremsin2002}. However, for our application and resolution requirements the assumption of a Gaussian distribution in both x- and y-direction can be justified. Corrections might be necessary at a later stage to reach the ultimate resolution. Currently, we apply the centroiding algorithm independently for the x- and y-direction in order to allow for different widths in the direction with the bias angle and perpendicular to it.

The driving requirements for our selection of the centroiding algorithm are to reach a high throughput that is not limiting the maximum event rate and to have a fixed run-through time per event. Furthermore, the algorithm needs to be implemented in binary logic on the FPGA of our readout electronics and thus has to work with integers without costly use of excessive mathematical operations. This ruled out the common solution in data analysis of a applying a least-squares fitting approach and leaves only non-iterative methods.

Such a method has been proposed by Caruana et al.\@ (1986)\cite{Caruana1986}. They used the fact that a Gaussian function is the exponent of a quadratic function and can be converted to a second-order polynomial by applying a logarithmic transformation:

Starting with a Gaussian distribution for the charge $z$ over the anode position $x$
\[
z(x) = A\exp{-\frac{(x-\overline{x})^2}{2\sigma^2}},
\]
and then applying the natural logarithm
\[
\ln(z) = \ln(A) + -\frac{(x-\overline{x})^2}{2\sigma^2} = 
\ln(A) - \frac{\overline{x}^2}{2\sigma^2} + \frac{\overline{x}}{\sigma^2}x - \frac{1}{2\sigma^2}x^2
\]
leads to a second order polynomial of the form
\[
\ln(z) = a + bx + cx^2.
\]
Its three parameters can be expressed with the parameters of the original Gaussian (total area $A$; central position $\overline{x}$; standard deviation $\sigma$):
\[
a = \ln(A) - \frac{\overline{x}^2}{2\sigma^2}, \hspace{.5cm}
b = \frac{\overline{x}}{\sigma^2}, \hspace{.5cm}
c = - \frac{1}{2\sigma^2}.
\]
A least-square approach requires minimizing
\[
E(x) = \left(\ln(z) - \left(a + bx + cx^2\right)\right)^2,
\]
which leads to a system of three linear equations
\[
\begin{bmatrix}
N & \sum_i x_i & \sum_i x_i^2 \\
\sum_i x_i & \sum_i x_i^2 & \sum_i x_i^3 \\
\sum_i x_i^2 & \sum_i x_i^3 & \sum_i x_i^4 \\
\end{bmatrix}
\begin{bmatrix}
a \\ b \\ c
\end{bmatrix}
=
\begin{bmatrix}
\sum_i \ln(z_i) \\ \sum_i \ln(z_i)x_i \\ \sum_i \ln(z_i)x_i^2
\end{bmatrix}
\]
with $N$ the number of available data points $(x_i,z_i)$, i.e.\@ the measured charge $z_i$ on the strip $x_i$.

However, Guo (2012) showed that this approach fails in determining the parameters $a$, $b$, and $c$ accurately when noise is present in the data and introduces a weight to the data to mitigate\cite{Guo2012}. The function $z(x)$ has to be modified to the function $\hat{z}(x)$ in order to account for the noise $\eta$ and represent the observed data:
\[
\hat{z}(x) = z(x) + \eta
\]
Conducting a Taylor expansion and omitting all but the zeroth order term leads to a modified error function that includes the square of the measured value $\hat{z}^2$ as weight:
\[
\hat{E}(x) = \hat{z}^2(x) \left[\ln(\hat{z}) - \left(a + bx + cx^2\right)\right]^2
\]
This again leads to a linear system of equations enlarged by the weights 
\[
\begin{bmatrix}
\sum_i \hat{z_i}^2 & \sum_i x_i\hat{z_i}^2 & \sum_i x_i^2\hat{z_i}^2 \\
\sum_i x_i\hat{z_i}^2 & \sum_i x_i^2\hat{z_i}^2 & \sum_i x_i^3\hat{z_i}^2 \\
\sum_i x_i^2\hat{z_i}^2 & \sum_i x_i^3\hat{z_i}^2 & \sum_i x_i^4\hat{z_i}^2 \\
\end{bmatrix}
\begin{bmatrix}
a \\ b \\ c
\end{bmatrix}
=
\begin{bmatrix}
\sum_i \hat{z_i}^2\ln(\hat{z}_i) \\ \sum_i \hat{z_i}^2\ln(\hat{z}_i)x_i \\ \sum_i \hat{z_i}^2\ln(\hat{z}_i)x_i^2
\end{bmatrix}.
\]
The solution of this system of equations yields the parameters b and c from which the position of the center of the Gaussian distribution can be calculated:
\[
\overline{x} = \frac{-b}{2c}.
\]

The full derivation of this approach can be found in Guo (2012)\cite{Guo2012}. Pastuchova and Zakopcan (2015) present a direct comparison of Caruana's original algorithm and the modification with Guo's proposed weighting solution by using experimental data \cite{Pastuchova2015}.

A similar but simpler method that also applies a logarithm transformation for centroiding the charge information for a cross strip anode was presented by Tremsin et al.\@ (2003) as "Gaussian 3-point centroiding"\cite{Tremsin2003}. However, this takes only the three strips with the highest charge value into account, while our implementation takes all strips above a defined threshold to determine the centroid position. This should lead to a higher accuracy with moderate additional needs for computing resources. Further more sophisticated comparisons with simulated and measurement data are planned.

\subsection{First Results}
\label{ssec:impl}
In order to save space in the FPGA and to speed up the centroiding calculations, a look-up table (LUT) is used for $\ln(\hat{z})\hat{z}^2$. For the x- and the y-axis the algorithm is exactly the same and in the FPGA implementation both axes are processed simultaneously in two parallel pipelines to yield independently the x- and y-position of an event.

\begin{figure} [ht]
  \begin{center}
    \includegraphics[width=\textwidth]{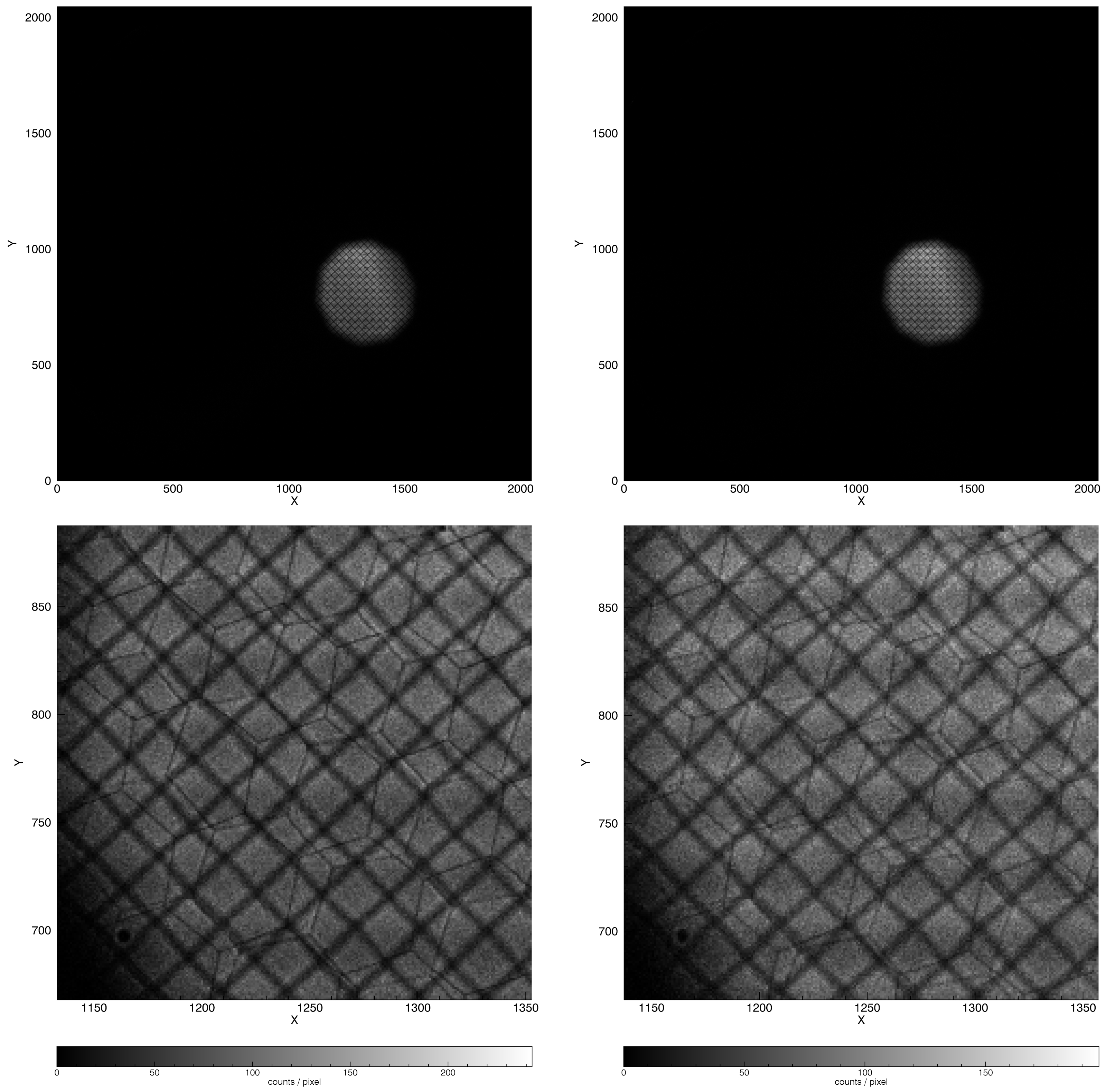}
   \end{center}
  \caption
  { \label{fig:comp}
  Image of an etched metal grid placed in front of the detector to compare centroiding methods: Gaussian fitting (left panels) and the non-iterative approach presented in Sect.~\ref{ssec:algo} (right panels). Both were implemented in software and run on pre-processed events taken with a detector prototype. While the top panels show the full detector images, the lower panels zoom on the illuminated region.
  The etched grid was tilted by 45° to detect possible non-linearities that could arise from the interpolation of the centroiding process between two anode strips. These slight nonlinearities can be seen as a small distortion of the grid and a faint dark and light pattern in the horizontal and vertical directions. (Coincidentally, the diagonal crossing points of the grid have nearly the same distance as the anode strips.) In both images the hexagonal substructure of the MCPs is clearly visible. However, the contrast is still slightly higher for Gaussian fitting, leaving room for optimization of the non-iterative approach.}
\end{figure}

We use conventional least-square fitting of Gaussians as a benchmark for any centroiding algorithm we test. Comparisons with the above described non-iterative centroiding have been performed first in software with recorded events for which the full raw data of all 128 strips have been read out and preprocessed in a computer. The results of this comparison between Gaussian fitting and the non-iterative algorithm are presented in Fig.~\ref{fig:comp}. While the sharpness of the image with fitted centroiding is still slightly higher, also the non-iterative centroiding preserves fine details. For example, pore crushing that occurs during the manufacturing process of the MCPs leads to slightly lower gain and sensitivity in the MCP channels located at hexagonal boundaries (see the lower panels of Fig.~\ref{fig:comp}). This usually unwanted effect (which can be corrected by an appropriate flat field calibration) helps here in determining the achieved resolution of the centroiding method used.

The mean deviation of the pixel position between the two methods is well below one pixel.

\subsection{Quantitative Results}

In order to allow direct testing and verification of the implementation of the non-iterative algorithm in the FPGA via a VHDL description, an FPGA firmware was developed. This firmware processes the same preprocessed events that were used in the software tests described above.

This approach allowed for a direct and quantitative comparison between the three discussed centroiding methods: Gaussian fit and the non-iterative algorithm in software and in the FPGA.

\begin{figure} [ht]
  \begin{center}
    \includegraphics[width=\textwidth]{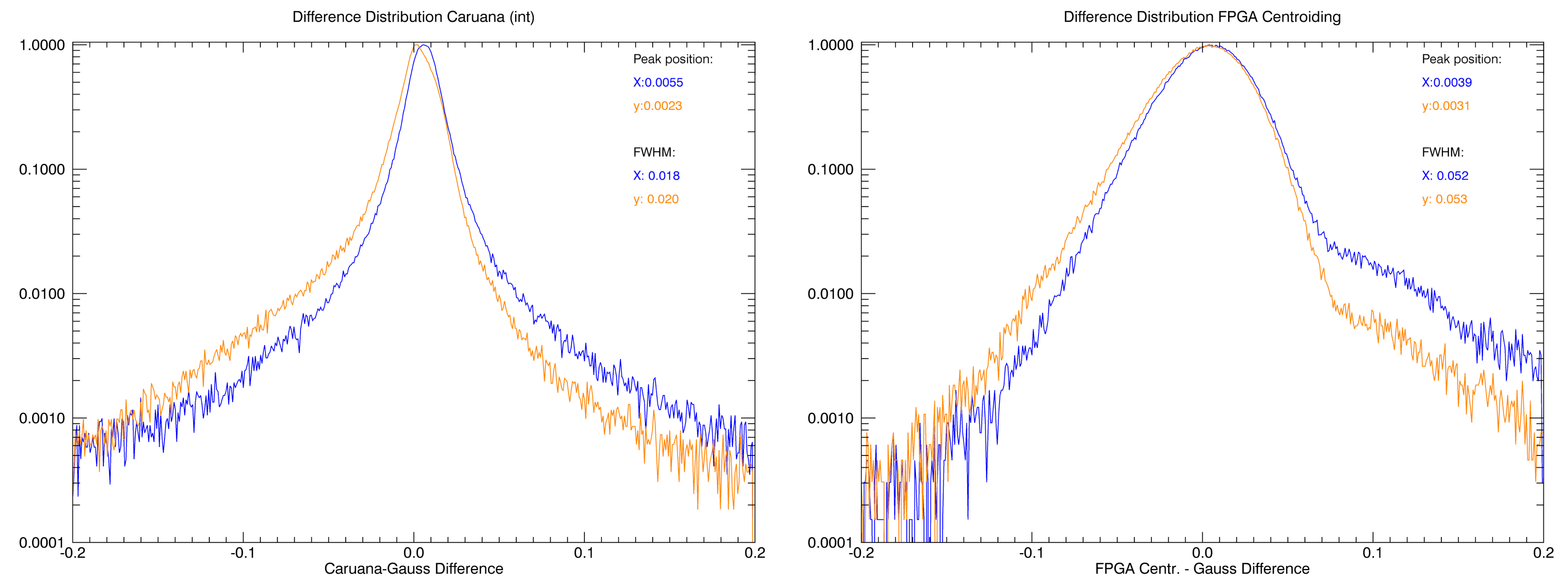}
   \end{center}
  \caption
  { \label{fig:DiffDistr} 
  Distribution of the position differences between the Gaussian fit positions and the non-iterative algorithm position, in software (left) and in the FPGA (right). Two curves are shown: one for the x positions (blue) and one for the y positions (orange). Units of the horizontal axis are the distance between two anode strips (=1.0). As we interpolate to 1/32 between two strips, one pixel width corresponds to about 0.03 on the horizontal axis. The inset lists the widths of the distributions as full width at half maximum (FWHM).}
\end{figure}

Fig.~\ref{fig:DiffDistr} shows the distribution of the differences of the positions between the Gaussian fit positions and the non-iterative algorithm positions, on the left side calculated with the software implementation and on the right side with the FPGA hardware implementation. Each diagram shows two curves, one for the x positions (blue) and one for the y positions (orange). The asymmetry between x and y curves may result from the bias angle of the MCPs, which leads to an asymmetric electron distribution in the electron cloud that hits the anode. The widths of these distributions are with FWHM$_x$ = 0.018 and FWHM$_y$ = 0.020 for the software implementation well below the size of one pixel of 0.03. For the FPGA hardware implementation the widths are FWHM$_x$ = 0.052 and FWHM$_y$ = 0.053 and therefore almost 2 pixel.

\begin{figure} [ht]
  \begin{center}
    \includegraphics[width=\textwidth]{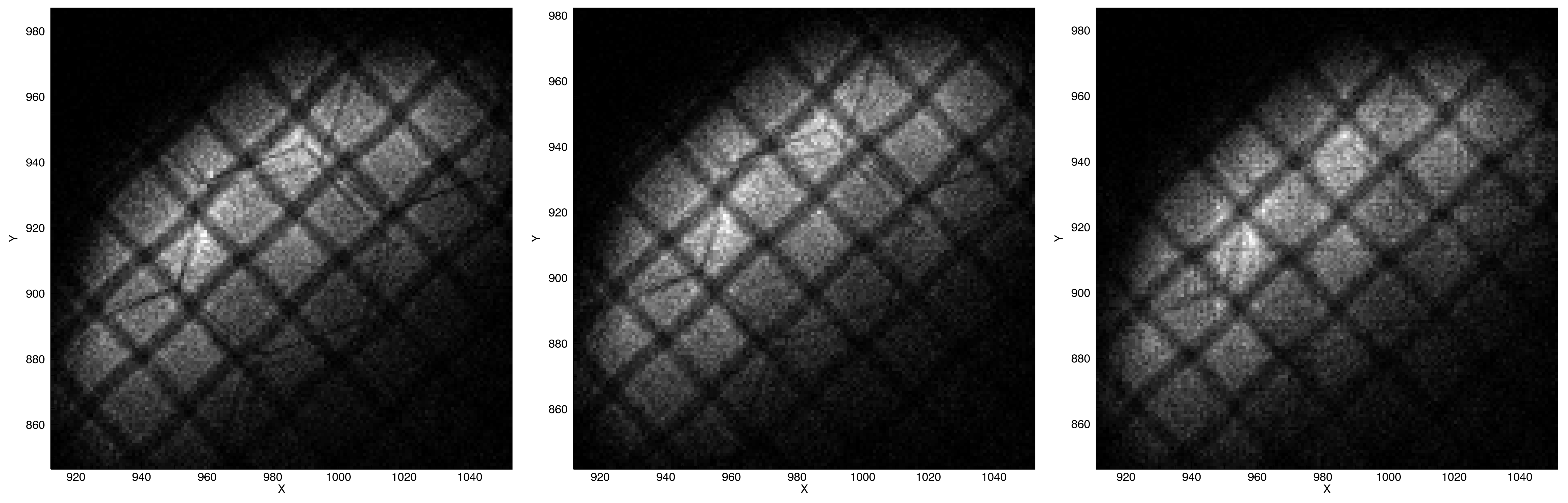}
   \end{center}
  \caption
  { \label{fig:comp_FPGA} 
Comparison of the resolutions achieved by centroiding with Gaussian fitting (left), with the integer algorithm implemented in software (center), and with the current implementation in the readout electronics FPGA (right). The reduced sensitivity and gain of MCP channels located at hexagonal boundaries creates contrast that effectively illustrates the spatial resolution of the detector. While Gaussian fitting and the software implementation produce almost equal results, the FPGA implementation currently seems to suffer an insufficient bit depth during processing.}
\end{figure}

Comparison images of Gaussian fitting and non-iterative centroiding in software and on the FPGA can be seen in Fig.~\ref{fig:comp_FPGA}. Slight artifacts such as horizontal and vertical structures are visible in the centroiding on the FPGA. This can be attributed to an insufficient bit depth in the FPGA pipeline, and will be solved in the next iteration of the implementation. 

The maximum integration count rate currently achieved with the FPGA hardware implementation is about 28,000\,cts/s. An optimization with consequent use of pipelining methods will be needed to achieve higher integration rates. Further optimization to decrease the required space on the FPGA for the pipelines is ongoing.

\subsection{Nonlinearity and its Correction}

\begin{figure}
  \begin{center}
    \includegraphics[width=.8\textwidth]{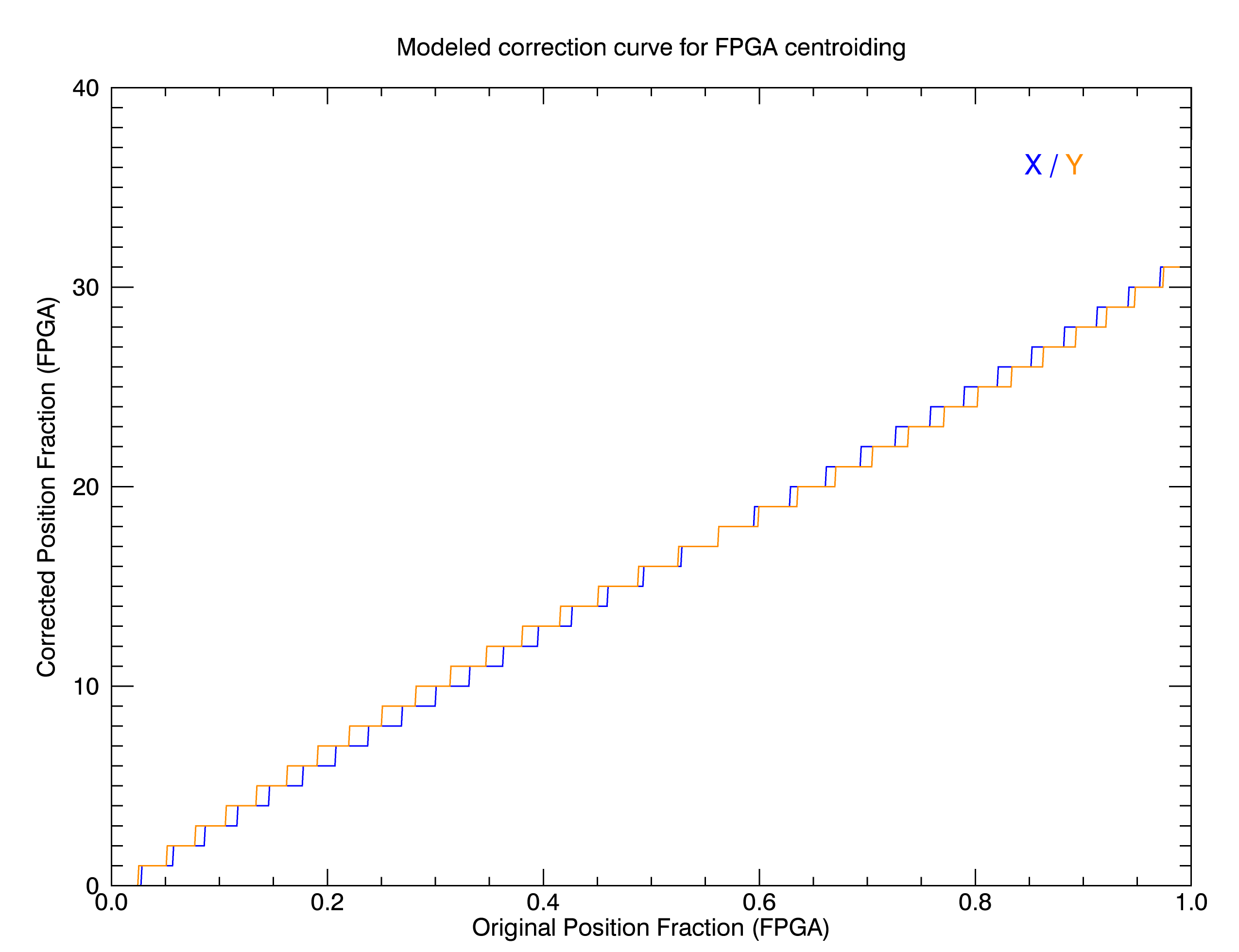}
  \end{center}
  \caption[modeled correction curve for FPGA centroiding]
  { \label{fig:CorrCurve_FPGA} 
  This is the modeled correction curve for the event position fraction values, i.e.\@ the interpolated values between 0.0 and 1.0 for the positions between two anode strips. These corrections are used for the FPGA hardware centroiding and implemented as lookup table. The horizontal axis shows the input fraction as 10 bit data (1024 values) ranging from 0.000 to about 0.999. The vertical axis shows the 5 bit output values as integers from 0 to 31. The correction has to be done for x and y separately with different values.}
\end{figure}

\begin{figure}
  \begin{center}
    \includegraphics[width=\textwidth]{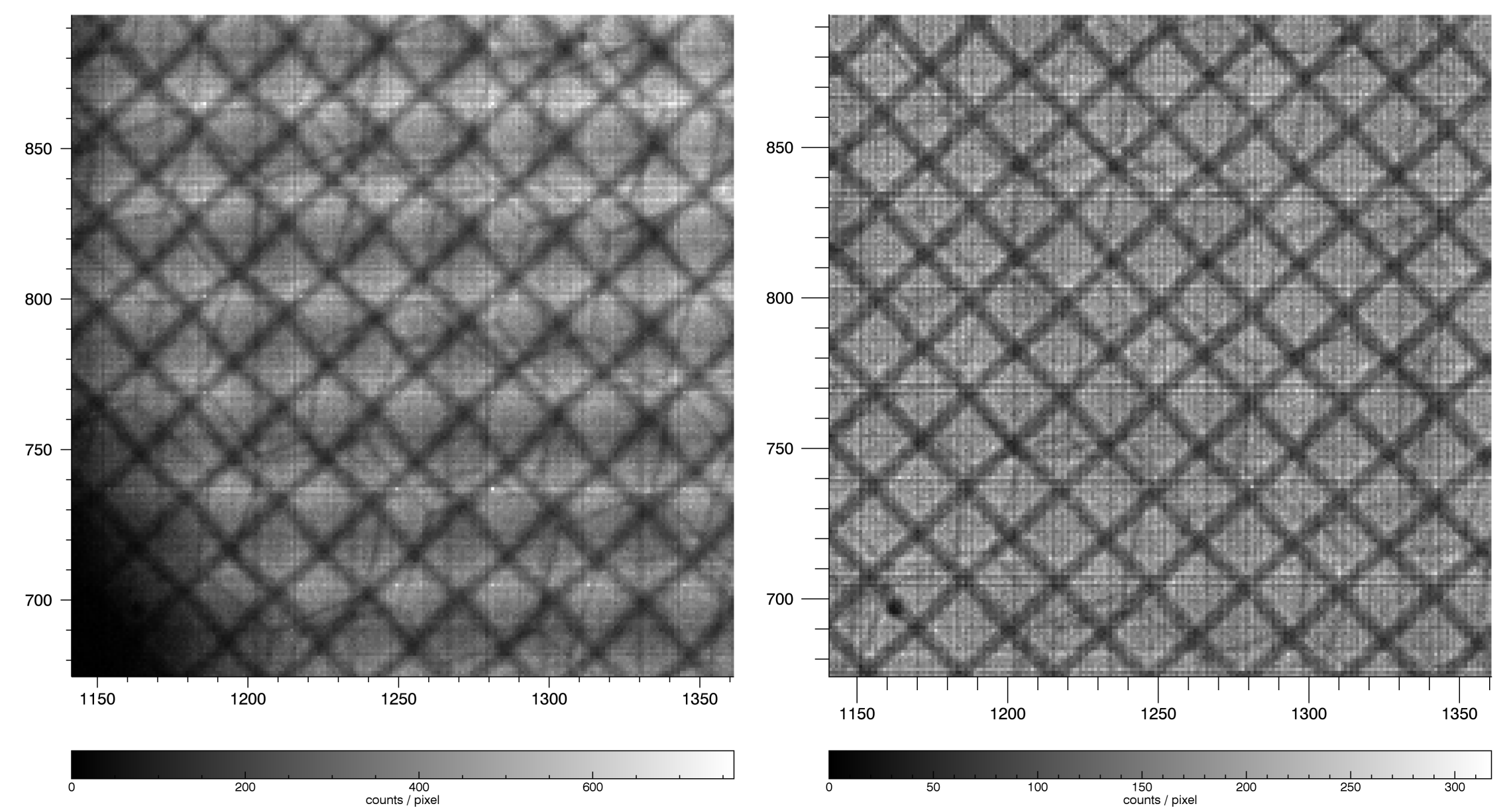}
   \end{center}
  \caption
  { \label{fig:FlatField_FPGA} 
  These images show integrations with the FPGA hardware implementation: Without nonlinearity correction (left) and with nonlinearity correction (right). The implemented nonlinearity correction nearly eliminates the weak periodic dark and light patterns as well as the distortions of the metal grid in front of the detector. (The darker area on the lower left-hand-side of the left image is due to shading and thus unrelated to the nonlinearity effect.)}
\end{figure}

Slight nonlinearities are caused by the centroiding process as we have a limited number of data points to estimate the center of the Gaussian function and also the distribution of the electron cloud hitting the anode is not perfectly Gaussian. The introduced nonlinearity leads to a periodic compression and stretching of the pixels between two anode strips which finally results in a local distortion of the image and also in weak dark and light areas caused by different pixel areas. Measuring the overall intensity variation for the position fractions 0.0 to 1.0 between two anode strips and assuming a flat field illumination one can easily reconstruct the distortion and model a correction curve. Fig.~\ref{fig:CorrCurve_FPGA} shows this modeled correction curve used for the FPGA hardware centroiding. It is implemented as 1024~$\times$~5~bit lookup-table for each direction (x and y). As the correction is a sub-pixel correction, the input data have to be of higher resolution than the output data. In the current version we use 10 bit data (1024 values) for the input fraction (the position between two anode strips: 0.0 to 0.999) and get the required 5~bit corrected fraction value. Fig.~\ref{fig:FlatField_FPGA} show images generated by the FPGA hardware centroiding without (left) and with (right) this position correction. It is clearly visible that the periodic darker and lighter areas are completely eliminated by the correction. Also the distortions of the metal grid in front of the detector are minimized so that the grid lines appear straight and sharper in the image.

\clearpage

\section{Conclusions}
\label{sect:conc}

Considerable efforts are undertaken at IAAT for the development of advanced MCP-based detector systems for UV astronomy. The goal is to provide a sealed as well as an open-face version to cover the full UV range from 380\,nm down to 10\,nm. Since a sealed prototype was successfully produced already some years ago, we currently focus on the open-face version with an opaque Al$_{x}$Ga$_{1-x}$N photocathode coated directly on the MCP stack. Both versions share the same coplanar cross-strip anode and the FPGA-based readout electronics.

In the first part of this publication the results of a study with different Al$_{x}$Ga$_{1-x}$N films on MgO with varied Al fraction $x$ are presented, in which a minimal lattice mismatch was found at a much lower Al fraction than initially expected. Furthermore, the measured bandgaps for the different compositions are reported. We will continue this systematic work by progressively investigating the growth of Al$_{x}$Ga$_{1-x}$N directly onto ALD-coated MgO layers, similar to those deposited on borosilicate MCPs.

In the second part, it has been shown that the presented non-iterative centroiding approach delivers promising results that almost reach the resolution of the benchmark obtained with Gaussian fitting. The new centroiding algorithm is already running on our FPGA-based readout electronics and further optimization of the implementation is in progress, e.g.\@ revising the bit depth of all pipeline steps and an increased use of LUTs. In parallel, we study how charge cloud asymmetry\cite{Tremsin2002} and subtle electronics effects\cite{Lapington1998} can be taken into account to further boost the centroiding resolution.

\subsection*{Disclosures}
The authors have no conflicts of interest to disclose concerning this manuscript.

\subsection* {Code, Data, and Materials Availability}
Regarding the Al$_{x}$Ga$_{1-x}$N photocathode development described in Sect.~\ref{sect:gan}, the main results are fully contained in Table~\ref{tab:latticeconst}. All other code and data, e.g.\@ concerning the characterization of substrates and films as well as the non-iterative centroiding method described in Sect.~\ref{sect:centroiding}, can be shared and discussed via e-mail with the corresponding author.

\subsection* {Acknowledgments}
This work was supported by the Bundesministerium für Wirtschaft und Klimaschutz through the Deutsches Zentrum für Luft- und Raumfahrt e.V.\@ (DLR) under the grant number 50 QT 2001. Contacts to samples for ongoing studies of p-doping in Al$_{x}$Ga$_{1-x}$N films were prepared in an evaporation system funded partially by the Deutsche Forschungsgesellschaft (DFG INST 189/191-1 FUGG).


\bibliography{2025-JATIS-MCP_}

@Article{Zhou17,
  author    = {Zhou, Chuanle and Ghods, Amirhossein and Saravade, Vishal G. and Patel, Paresh V. and Yunghans, Kelcy L. and Ferguson, Cameron and Feng, Yining and Kucukgok, Bahadir and Lu, Na and Ferguson, Ian T.},
  journal   = {ECS Journal of Solid State Science and Technology},
  title     = {{Review—The Current and Emerging Applications of the III-Nitrides}},
  year      = {2017},
  month     = {nov},
  number    = {12},
  pages     = {Q149},
  volume    = {6},
  doi       = {10.1149/2.0101712jss},
  publisher = {The Electrochemical Society},
  url       = {https://dx.doi.org/10.1149/2.0101712jss},
}

@InProceedings{Bayr17,
  author    = {Bayram, C. and Liu, R.},
  booktitle = {Quantum Sensing and Nano Electronics and Photonics XIV},
  title     = {Polarization-free integrated gallium-nitride photonics},
  year      = {2017},
  editor    = {Razeghi, Manijeh},
  month     = jan,
  pages     = {101110Y},
  publisher = {SPIE},
  volume    = {10111},
  doi       = {10.1117/12.2251607},
  issn      = {0277-786X},
}

@Article{Park00,
  author    = {Park, Seoung-Hwan and Chuang, Shun-Lien},
  journal   = {Journal of Applied Physics},
  title     = {{Comparison of zinc-blende and wurtzite GaN semiconductors with spontaneous polarization and piezoelectric field effects}},
  year      = {2000},
  issn      = {1089-7550},
  month     = jan,
  number    = {1},
  pages     = {353--364},
  volume    = {87},
  doi       = {10.1063/1.371915},
  publisher = {AIP Publishing},
}

@Article{Kaku13,
  author    = {Kakuda, M. and Morikawa, S. and Kuboya, S. and Katayama, R. and Yaguchi, H. and Onabe, K.},
  journal   = {Journal of Crystal Growth},
  title     = {{RF-MBE growth of cubic AlN on MgO (001) substrates via 2-step c-GaN buffer layer}},
  year      = {2013},
  issn      = {0022-0248},
  month     = sep,
  pages     = {307--309},
  volume    = {378},
  doi       = {10.1016/j.jcrysgro.2012.12.120},
  publisher = {Elsevier BV},
}

@Article{Zhan07,
  author    = {Zhang, Xinyu and Chen, Zhouwen and Zhang, Shiliang and Liu, Riping and Zong, Haitao and Jing, Qin and Li, Gong and Ma, Mingzhen and Wang, Wenkui},
  journal   = {Journal of Physics: Condensed Matter},
  title     = {Electronic and optical properties of rock-salt aluminum nitride obtained from first principles},
  year      = {2007},
  issn      = {1361-648X},
  month     = sep,
  number    = {42},
  pages     = {425231},
  volume    = {19},
  doi       = {10.1088/0953-8984/19/42/425231},
  publisher = {IOP Publishing},
}

@Article{Vega21,
  author    = {Vegard, L.},
  journal   = {Zeitschrift für Physik},
  title     = {{Die Konstitution der Mischkristalle und die Raumfüllung der Atome}},
  year      = {1921},
  issn      = {1434-601X},
  month     = jan,
  number    = {1},
  pages     = {17--26},
  volume    = {5},
  doi       = {10.1007/bf01349680},
  publisher = {Springer Science and Business Media LLC},
}

@Article{Maku18,
  author    = {Makuła, Patrycja and Pacia, Michał and Macyk, Wojciech},
  journal   = {The Journal of Physical Chemistry Letters},
  title     = {{How To Correctly Determine the Band Gap Energy of Modified Semiconductor Photocatalysts Based on UV–Vis Spectra}},
  year      = {2018},
  issn      = {1948-7185},
  month     = dec,
  number    = {23},
  pages     = {6814--6817},
  volume    = {9},
  doi       = {10.1021/acs.jpclett.8b02892},
  publisher = {American Chemical Society (ACS)},
}

@Article{Adac85,
  author    = {Adachi, Sadao},
  journal   = {Journal of Applied Physics},
  title     = {{GaAs, AlAs, and AlxGa1-xAs: Material parameters for use in research and device applications}},
  year      = {1985},
  issn      = {1089-7550},
  month     = aug,
  number    = {3},
  pages     = {R1--R29},
  volume    = {58},
  doi       = {10.1063/1.336070},
  publisher = {AIP Publishing},
}

@Article{Gonz07,
  author    = {González, Rafael and López, William and Rodríguez M., Jairo Arbey},
  journal   = {Solid State Communications},
  title     = {{First-principles calculations of structural properties of GaN: V}},
  year      = {2007},
  issn      = {0038-1098},
  month     = oct,
  number    = {3–4},
  pages     = {109--113},
  volume    = {144},
  doi       = {10.1016/j.ssc.2007.08.024},
  publisher = {Elsevier BV},
}

@Article{Fu12,
  author    = {Fu, Yuechun and Zhang, Yao and Yang, Weijia and He, Huan and Shen, Xiaoming},
  journal   = {Journal of Crystal Growth},
  title     = {{Surface evolution of NaCl-type cubic AlN films on MgO (100) substrates deposited by laser molecular beam epitaxy}},
  year      = {2012},
  issn      = {0022-0248},
  month     = mar,
  number    = {1},
  pages     = {28--32},
  volume    = {343},
  doi       = {10.1016/j.jcrysgro.2012.01.030},
  publisher = {Elsevier BV},
}

@Article{Pank02,
  author    = {Pankov, V. and Evstigneev, M. and Prince, R. H.},
  journal   = {Applied Physics Letters},
  title     = {{Role of substrate in the pseudomorphic stabilization of rocksalt-type AlN phase in AlN/TiN superlattices}},
  year      = {2002},
  issn      = {1077-3118},
  month     = jun,
  number    = {22},
  pages     = {4142--4144},
  volume    = {80},
  doi       = {10.1063/1.1482798},
  publisher = {AIP Publishing},
}

@Article{Tremsin1999,
  author    = {Tremsin, A. S. and Siegmund, O. H. W.},
  journal   = {Review of Scientific Instruments},
  title     = {{Spatial distribution of electron cloud footprints from microchannel plates: Measurements and modeling}},
  year      = {1999},
  issn      = {1089-7623},
  month     = aug,
  number    = {8},
  pages     = {3282--3288},
  volume    = {70},
  doi       = {10.1063/1.1149905},
  publisher = {AIP Publishing},
}

@Article{Caruana1986,
  author    = {Caruana, Richard A. and Searle, Roger B. and Heller, Thomas. and Shupack, Saul I.},
  journal   = {Analytical Chemistry},
  title     = {Fast algorithm for the resolution of spectra},
  year      = {1986},
  issn      = {1520-6882},
  month     = may,
  number    = {6},
  pages     = {1162--1167},
  volume    = {58},
  doi       = {10.1021/ac00297a041},
  publisher = {American Chemical Society (ACS)},
}

@Misc{Guo2012,
  author    = {Guo, Hongwei},
  journal   = {Streamlining Digital Signal Processing},
  month     = jun,
  title     = {A Simple Algorithm for Fitting a Gaussian Function},
  year      = {2012},
  doi       = {10.1002/9781118316948.ch31},
  isbn      = {9781118316948},
  pages     = {297--305},
  publisher = {Wiley},
}

@Article{Tremsin2020,
  author    = {Tremsin, A.S. and Vallerga, J.V.},
  journal   = {Radiation Measurements},
  title     = {{Unique capabilities and applications of Microchannel Plate (MCP) detectors with Medipix/Timepix readout}},
  year      = {2020},
  issn      = {1350-4487},
  month     = jan,
  pages     = {106228},
  volume    = {130},
  doi       = {10.1016/j.radmeas.2019.106228},
  publisher = {Elsevier BV},
}

@Article{Siegmund2003,
  author    = {Siegmund, O.H.W. and Tremsin, A.S. and Vallerga, J.V. and Abiad, R. and Hull, J.},
  journal   = {Nuclear Instruments and Methods in Physics Research Section A: Accelerators, Spectrometers, Detectors and Associated Equipment},
  title     = {{High resolution cross strip anodes for photon counting detectors}},
  year      = {2003},
  issn      = {0168-9002},
  month     = may,
  number    = {1–3},
  pages     = {177--181},
  volume    = {504},
  doi       = {10.1016/s0168-9002(03)00816-7},
  publisher = {Elsevier BV},
}

@InProceedings{Tremsin2002,
  author    = {Tremsin, Anton S. and Siegmund, Oswald H. W.},
  booktitle = {X-Ray and Gamma-Ray Instrumentation for Astronomy XII},
  title     = {{Charge cloud asymmetry in detectors with biased MCPs}},
  year      = {2002},
  editor    = {Flanagan, Kathryn A. and Siegmund, Oswald H. W.},
  month     = jan,
  pages     = {127--138},
  publisher = {SPIE},
  volume    = {4497},
  doi       = {10.1117/12.454218},
  issn      = {0277-786X},
}

@InProceedings{Lapington1998,
  author    = {Lapington, Jonathan S. and Sanderson, B. S. and Worth, Liam B. C.},
  booktitle = {EUV, X-Ray, and Gamma-Ray Instrumentation for Astronomy IX},
  title     = {{The Vernier electronic readout: high resolution and image stability from a charge division readout for microchannel plates}},
  year      = {1998},
  editor    = {Siegmund, Oswald H. W. and Gummin, Mark A.},
  month     = nov,
  pages     = {535--545},
  publisher = {SPIE},
  volume    = {3445},
  doi       = {10.1117/12.330316},
  issn      = {0277-786X},
}

@InProceedings{Tremsin2003,
  author    = {Tremsin, Anton S. and Vallerga, John V. and Siegmund, Oswald H. W. and Hull, Jeff S.},
  booktitle = {UV/EUV and Visible Space Instrumentation for Astronomy II},
  title     = {{Centroiding algorithms and spatial resolution of photon counting detectors with cross-strip anodes}},
  year      = {2003},
  editor    = {Siegmund, Oswald H. W.},
  month     = dec,
  pages     = {113},
  publisher = {SPIE},
  volume    = {5164},
  doi       = {10.1117/12.508409},
  issn      = {0277-786X},
}

@InBook{Diebold2022,
  author    = {Diebold, Sebastian},
  editor    = {Bambi, C. and Santangelo, A.},
  pages     = {1--36},
  publisher = {Springer Nature Singapore},
  title     = {{Proportional Counters and Microchannel Plates}},
  year      = {2022},
  isbn      = {9789811645440},
  month     = nov,
  booktitle = {Handbook of X-ray and Gamma-ray Astrophysics},
  doi       = {10.1007/978-981-16-4544-0_16-1},
}

@Article{Lapington2002,
  author    = {Lapington, J.S and Sanderson, B and Worth, L.B.C and Tandy, J.A},
  journal   = {Nuclear Instruments and Methods in Physics Research Section A: Accelerators, Spectrometers, Detectors and Associated Equipment},
  title     = {{Imaging achievements with the Vernier readout}},
  year      = {2002},
  issn      = {0168-9002},
  month     = jan,
  number    = {1–3},
  pages     = {250--255},
  volume    = {477},
  doi       = {10.1016/s0168-9002(01)01840-x},
  publisher = {Elsevier BV},
}

@Article{Pastuchova2015,
  author    = {Pastuchová, Elena and Zákopčan, Michal},
  journal   = {Journal of Electrical Engineering},
  title     = {{Comparison of Algorithms For Fitting a Gaussian Function Used in Testing Smart Sensors}},
  year      = {2015},
  issn      = {1339-309X},
  month     = may,
  number    = {3},
  pages     = {178--181},
  volume    = {66},
  doi       = {10.2478/jee-2015-0029},
  publisher = {Walter de Gruyter GmbH},
}

@InProceedings{Diebold2024,
  author    = {Diebold, Sebastian J. and Barnstedt, Jürgen and Bluhm, Jonas and Conti, Lauro and Elsener, Hans R. and Höltzli, Markus and Kalkuhl, Christoph and Rau, Darleen and Rupp, Laurin and Schaadt, Daniel and Schanz, Thomas and Stelzer, Beate and Stock, Alexander and Weinert, Christopher and Werner, Klaus and Wildfang, Tim},
  booktitle = {Space Telescopes and Instrumentation 2024: Ultraviolet to Gamma Ray},
  title     = {{Microchannel plate detectors for ultraviolet astronomy}},
  year      = {2024},
  editor    = {den Herder, Jan-Willem A. and Nakazawa, Kazuhiro and Nikzad, Shouleh},
  month     = aug,
  pages     = {25},
  publisher = {SPIE},
  doi       = {10.1117/12.3021533},
}

@InProceedings{Conti2022,
  author    = {Conti, Lauro and Barnstedt, Jürgen and Diebold, Sebastian J. and Höltzli, Markus and Kalkuhl, Christoph and Kappelmann, Norbert and Rauch, Thomas and Schanz, Thomas and Stelzer, Beate and Stock, Alexander and Werner, Klaus and Elsener, Hans-Rudolf and Meyer, Kevin and Schaadt, Daniel},
  booktitle = {Space Telescopes and Instrumentation 2022: Ultraviolet to Gamma Ray},
  title     = {{A photon counting imaging detector for UV space missions}},
  year      = {2022},
  editor    = {den Herder, Jan-Willem A. and Nakazawa, Kazuhiro and Nikzad, Shouleh},
  month     = aug,
  publisher = {SPIE},
  doi       = {10.1117/12.2628735},
}

@Article{Conti2018,
  author    = {Conti, Lauro and Barnstedt, Jürgen and Hanke, Lars and Kalkuhl, Christoph and Kappelmann, Norbert and Rauch, Thomas and Stelzer, Beate and Werner, Klaus and Elsener, Hans-Rudolf and Schaadt, Daniel M.},
  journal   = {Astrophysics and Space Science},
  title     = {{MCP detector development for UV space missions}},
  year      = {2018},
  issn      = {1572-946X},
  month     = mar,
  number    = {4},
  volume    = {363},
  doi       = {10.1007/s10509-018-3283-4},
  publisher = {Springer Science and Business Media LLC},
}

@InProceedings{Ji2020,
  author    = {Ji, Li and Lou, Zheng and Zhang, Jinlong and Qiu, Keqiang and Li, Shuangying and Sun, Wei and Yan, Shuping and Zhang, Shuinai and Qian, Yuan and Wang, Sen and Werner, Klaus and Fang, Taotao and Wang, Tinggui and Barnstedt, Jürgen and Buntrock, Sebastian and Cai, Mingsheng and Chen, Wen and Conti, Lauro and Deng, Lei and Diebold, Sebastian and Fu, Shaojun and Guo, Jianhua and Hanke, Lars and Hong, Yilin and Kalkuhl, Christoph and Kappelmann, Norbert and Kaufmann, Thomas and Lei, Shijun and Li, Fu and Li, Xinfeng and Liu, Wei and Meyer, Kevin and Rauch, Thomas and Ruan, Ping and Schaadt, Daniel M. and Schanz, Thomas and Song, Qian and Stelzer, Beate and Wang, Zhanshan and Yang, Jianfeng and Zhang, Wei},
  booktitle = {Space Telescopes and Instrumentation 2020: Ultraviolet to Gamma Ray},
  title     = {{Mapping Lyman UV emission for diffuse sources}},
  year      = {2020},
  editor    = {den Herder, Jan-Willem A. and Nakazawa, Kazuhiro and Nikzad, Shouleh},
  month     = dec,
  pages     = {4},
  publisher = {SPIE},
  doi       = {10.1117/12.2561839},
}

@InProceedings{Diebold2022a,
  author    = {Diebold, Sebastian J. and Barnstedt, Jürgen and Chandra, Bharat and Conti, Lauro and Ghatul, Shubham and Kappelmann, Norbert and Mohan, Rekhesh and Murthy, Jayant and Nair, Binukumar G. and Prabha, Shanti and Rai, Richa and Safonova, Margarita and Stelzer, Beate and Werner, Klaus},
  booktitle = {Space Telescopes and Instrumentation 2022: Ultraviolet to Gamma Ray},
  title     = {{TINI -- a mission for FUV spectroscopy of extended objects}},
  year      = {2022},
  editor    = {den Herder, Jan-Willem A. and Nakazawa, Kazuhiro and Nikzad, Shouleh},
  month     = aug,
  pages     = {112},
  publisher = {SPIE},
  doi       = {10.1117/12.2630121},
}

@Article{Barstow2014,
  author    = {Barstow, M.A. and Casewell, S.L. and Holberg, J.B. and Kowalski, M.P.},
  journal   = {Advances in Space Research},
  title     = {{The status and future of EUV astronomy}},
  year      = {2014},
  issn      = {0273-1177},
  month     = mar,
  number    = {6},
  pages     = {1003--1013},
  volume    = {53},
  doi       = {10.1016/j.asr.2013.08.007},
  publisher = {Elsevier BV},
}

@Article{Tsai2019,
  author    = {Tsai, Y.-C. and Bayram, C.},
  journal   = {Science Reports},
  title     = {{Structural and Electronic Properties of Hexagonal and Cubic Phase AlGaInN Alloys Investigated Using First Principles Calculations}},
  year      = {2019},
  month     = apr,
  volume    = {9},
  doi       = {10.1038/s41598-019-43113-w},
  publisher = {nature},
}

@Article{Kanoun2005,
  author    = {Kanoun, M. B. and Goumri-Said, S. and Merad, A. E. and Mariette, H.},
  journal   = {Journal of Applied Physics},
  title     = {{Ab initio study of structural parameters and gap bowing in zinc-blende $Al_{x}Ga_{1-x}N$ and $Al_{x}In_{1-x}N$ alloys}},
  year      = {2005},
  issn      = {6},
  month     = sep,
  volume    = {98},
  doi       = {10.1063/1.2060931},
  publisher = {AIP Publishing},
}

@Article{Melton1969,
  author    = {Melton, W. A. and Pankove, J. I},
  journal   = {Journal of Crystal Growth},
  title     = {{GaN growth on sapphire}},
  year      = {1969},
  issn      = {1-2},
  month     = jun,
  pages     = {168--173},
  volume    = {178},
  doi       = {10.1016/S0022-0248(97)00082-1},
  publisher = {Elsevier BV},
}

@Article{Jana2014,
  author    = {Jana, S. K. and Mukhopadhyay, P and Ghosh, S. and Kabi, S. and Bag, A. and Kumar, R. and Biswas, D.},
  journal   = {Journal of Applied Physics},
  title     = {{High-resolution X-ray diffraction analysis of $Al_xGa{1-x}N/In_xGa{1-x}N/GaN$ on sapphire multilayer structures: Theoretical, simulations, and experimental observations}},
  year      = {2014},
  month     = may,
  volume    = {115},
  doi       = {10.1063/1.4875382},
  publisher = {AIP Publishing},
}

@Article{Nepal2005,
  author    = {Nepal, N. and Li, J. and Nakarmi, L. and Lin, J. Y. and Jiang, H. X.},
  journal   = {Applied Physics Letters},
  title     = {{Temperature and compositional dependence of the energy band gap of AlGaN alloys}},
  year      = {2005},
  issn      = {24},
  month     = dec,
  volume    = {87},
  doi       = {10.1063/1.2142333},
  publisher = {AIP Publishing},
}

@InProceedings{Ertley2018,
  author    = {Camden Ertley and Oswald Siegmund and John Vallerga and Anton Tremsin and Nathan Darling and Jeff Hull and Joe Tedesco and Travis Curtis and Catriana Paw U},
  booktitle = {Space Telescopes and Instrumentation 2018: Ultraviolet to Gamma Ray},
  title     = {Microchannel plate detectors for future {NASA} {UV} observatories},
  year      = {2018},
  editor    = {Jan-Willem A. den Herder and Kazuhiro Nakazawa and Shouleh Nikzad},
  month     = {7},
  publisher = {{SPIE}},
  doi       = {10.1117/12.2314197},
}

@Article{Muth1999,
  author    = {Muth, J. F. and Brown, J. D. and Johnson, M. A. L. and Yu, Zhonghai and Kolbas, R. M. and Cook, J. W. and Schetzina, J. F.},
  journal   = {MRS Internet Journal of Nitride Semiconductor Research},
  title     = {Absorption Coefficient and Refractive Index of GaN, AlN and AlGaN Alloys},
  year      = {1999},
  issn      = {1092-5783},
  number    = {S1},
  pages     = {502--507},
  volume    = {4},
  doi       = {10.1557/s1092578300002957},
  publisher = {Springer Science and Business Media LLC},
}

@Article{Curtis2025,
  author    = {Curtis, Travis and Tremsin, Anton and Siegmund, Oswald and McPhate, Jason and Nell, Nicholas and Tercero Macua, Dennis},
  journal   = {Journal of Astronomical Telescopes, Instruments, and Systems},
  title     = {MCP detectors: overview, advances, and prospects for Habitable Worlds Observatory},
  year      = {2025},
  issn      = {2329-4124},
  month     = jun,
  number    = {04},
  volume    = {11},
  doi       = {10.1117/1.jatis.11.4.042206},
  publisher = {SPIE-Intl Soc Optical Eng},
}

@InProceedings{Scowen2025,
  author       = {Scowen, Paul A and Bolcar, Matthew R and Zhao, Feng},
  title        = {{Technology Roadmapping and Maturation Planning to Enable UV-Vis Astrophysics with the Habitable Worlds Observatory}},
  year         = {2025},
  organization = {to be published in Proc. SPIE},
  url          = {https://ntrs.nasa.gov/citations/20250007324},
}
\bibliographystyle{spiejour}

\end{document}